\documentclass[conference]{IEEEtran}
\IEEEoverridecommandlockouts

\usepackage[noadjust]{cite}
\usepackage{amsmath,amssymb,amsfonts}
\usepackage{algorithmic}
\usepackage{graphicx}
\usepackage{subfig}
\usepackage{textcomp}
\usepackage{xcolor}
\usepackage{bm}

\def\BibTeX{{\rm B\kern-.05em{\sc i\kern-.025em b}\kern-.08em
    T\kern-.1667em\lower.7ex\hbox{E}\kern-.125emX}}
\begin{document}

\title{Multiplexing More Streams in the MU-MISO Downlink by Interference Exploitation Precoding
}

\author{\IEEEauthorblockN{Ang Li}
\IEEEauthorblockA{\textit{School of Electrical and Inf. Eng.} \\
\textit{University of Sydney}\\
Sydney, Australia \\
ang.li2@sydney.edu.au}
\and
\IEEEauthorblockN{Christos Masouros}
\IEEEauthorblockA{\textit{Dept. of Electronic and Electrical Eng.} \\
\textit{University College London}\\
London, U.K. \\
c.masouros@ucl.ac.uk}
\and
\IEEEauthorblockN{Yonghui Li, and Branka Vucetic}
\IEEEauthorblockA{\textit{School of Electrical and Inf. Eng.} \\
\textit{University of Sydney}\\
Sydney, Australia \\
\{yonghui.li, branka.vucetic\}@sydney.edu.au}
}

\maketitle

\begin{abstract}
In this paper, we study the interference exploitation precoding for the scenario where the number of streams simultaneously transmitted by the base station (BS) is larger than that of transmit antennas at the BS, and derive the optimal precoding structure by employing the pseudo inverse. We show that the optimal pre-scaling vector is equal to a linear combination of the right singular vectors that correspond to zero singular values of the coefficient matrix. By formulating the dual problem, the optimal precoding matrix can be expressed as a function of the dual variables in a closed form, and an equivalent quadratic programming (QP) formulation is further derived for computational complexity reduction. Numerical results validate our analysis and demonstrate significant performance improvements for interference exploitation precoding for the considered scenario.
\end{abstract}


\section{Introduction}
Multi-antenna wireless communication systems have received extensive research attention due to their significant performance gains over single-antenna systems, where precoding has been widely acknowledged as a promising application \cite{r1}. When the channel state information (CSI) is available at the transmitter side, precoding is able to support data transmission to multiple users simultaneously. Well-known precoding approaches include theoretically capacity-achieving dirty paper coding (DPC) \cite{r2}, non-linear precoding such as Tomlinson-Harashima precoding (THP) \cite{r3} and vector perturbation (VP) precoding \cite{r4}, and low-complexity linear precoding such as zero-forcing (ZF) and regularized ZF (RZF) \cite{r6}. Meanwhile, downlink precoding based on optimization has also received increasing research attention in recent years \cite{r7}\nocite{r10}-\cite{r11}. Among optimization-based precoding approaches, signal-to-interference-plus-noise ratio (SINR) balancing \cite{r10} and power minimization \cite{r11} are two most popular designs. These precoding schemes exploit the information of the channel to design the precoding matrices that target at avoiding or limiting the interference.

Compared to the above studies that treat interference as detrimental, recent studies show that interference can also be beneficial and provide further performance improvements on a symbol level \cite{r14}. By exploiting the information of the data symbols and their corresponding constellation, the instantaneous interference can be divided into constructive interference (CI) and destructive interference \cite{r15}. More specifically, CI is defined as the interference that pushes the received signals away from the detection thresholds, which further improves the detection performance. Based on the above, CI-based precoding for PSK modulations has been proposed in \cite{r16}, \cite{r17} as a modification of ZF precoding. Optimization-based CI precoding has further been proposed in \cite{r31} based on symbol scaling and \cite{r18}\nocite{r19}-\cite{r20} based on phase rotation, and their extension to multi-level modulations such as QAM is discussed in \cite{r21}, \cite{r33}. More recently, it has been revealed in \cite{r45} that there exists an optimal precoding structure for CI precoding. In addition to the performance improvements of CI precoding over traditional precoding approaches, another advantage of CI precoding is its capability of supporting a larger number of streams (single-antenna users) than the number of transmit antennas at the base station (BS) simultaneously, which has only been numerically shown in \cite{r18}. Nevertheless, it is still not clear whether the analysis and results in \cite{r45} can be extended to this scenario.

Therefore in this paper, we focus on the scenario where the number of streams simultaneously transmitted by the BS is larger than that of transmit antennas at the BS. Based on the Karush-Kuhn-Tucker (KKT) conditions, we derive the optimal precoding structure at the BS and transform the problem into an optimization on the pre-scaling vector. In the derivation, the exact matrix inverse is not applicable due to the rank deficiency, and accordingly we employ the pseudo inverse instead, which introduces an additional constraint to the optimization problem. We further show that the optimal pre-scaling vector is equal to a linear combination of the right singular vectors corresponding to zero singular values of the coefficient matrix. Subsequently, the optimization problem is further transformed into an optimization on the weights for each singular vector, which is finally shown to be a quadratic programming (QP) optimization and can be more efficiently solved than the original second-order cone programming (SOCP) formulation. Based on the QP formulation, we also discuss the condition under which multiplexing more streams than the number of transmit antennas at the BS is feasible with interference exploitation precoding. Numerical results validate our analysis and demonstrate significant performance gains of interference exploitation precoding over traditional precoding methods in the considered scenario.

{\it Notations:} $a$, $\bf a$, and $\bf A$ denote scalar, column vector and matrix, respectively. ${( \cdot )^*}$, ${( \cdot )^T}$, ${( \cdot )^H}$, ${( \cdot )^{-1}}$, and ${( \cdot )^+}$ denote conjugate, transposition, conjugate transposition, inverse, and pseudo inverse of a matrix, respectively. $diag \left(  \cdot  \right)$ is the transformation of a column vector into a diagonal matrix, and ${\rm{vec}}\left(  \cdot  \right)$ denotes the vectorization operation. $\left|  \cdot  \right|$ denotes the absolute value or the modulus, and $\left\|  \cdot  \right\|_2$ is the $l$2-norm. ${{\cal C}^{n \times n}}$ and ${{\cal R}^{n \times n}}$ represent an $n \times n$ matrix in the complex and real set, respectively. $j$ is the imaginary unit. 

\section{System Model and Problem Formulation}
\subsection{System Model}
We focus on a multi-user multiple-input single-ouput (MU-MISO) system in the downlink, where the BS with $N_t$ transmit antennas transmits a total number of $K$ streams simultanesouly, and $K>N_t$. The data symbol vector ${\bf s}=\left[ {s_1, s_2, \cdots, s_K} \right]^T \in {\cal C}^{K \times 1}$ is assumed to be from a normalized PSK constellation, and therefore $s_k s_k^H=1$ for each stream $k$ \cite{r18}. The received signal at user $k$ can be expressed as
\begin{equation}
r_k={\bf h}_k^T {\bf Ws} + n_k,
\label{eq_1}
\end{equation}
where ${\bf h}_k \in {\cal C}^{N_t \times 1}$ denotes the flat-fading Rayleigh channel between user $k$ and the BS, and perfect CSI is assumed throughout this paper. ${\bf W} \in {\cal C}^{N_t \times K}$ is the precoding matrix, and $n_k$ is the additive Gaussian nose with zero mean and variance $\sigma^2$.

\begin{figure}[!b]
\centering
\includegraphics[scale=0.32]{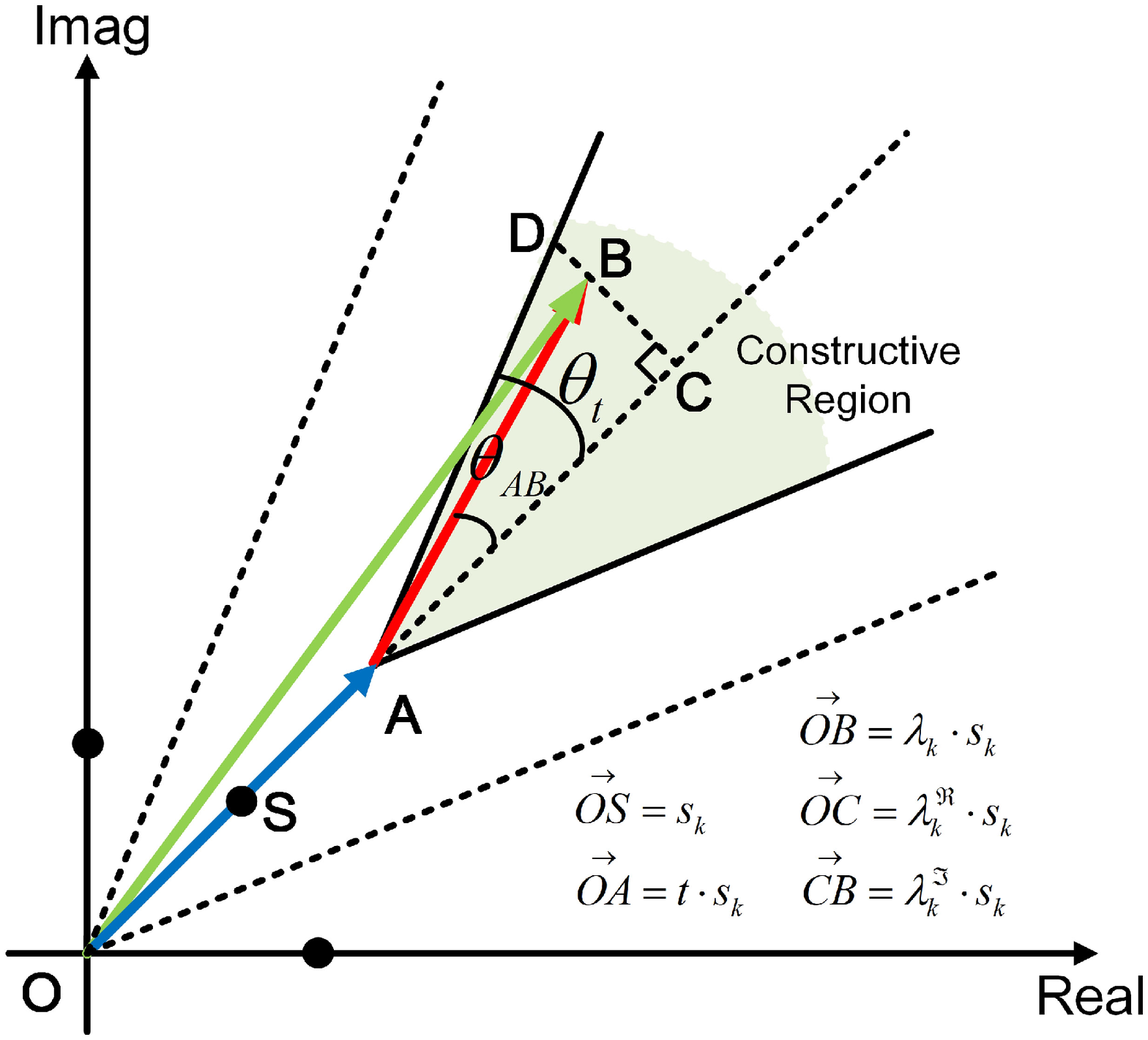}
\caption{The phase-rotation metric for 8PSK constellation}
\end{figure}

\subsection{Problem Formulation}
The optimization of CI precoding can be formulated based on the geometry of the PSK constellation, as shown in Fig. 1 where we employ 8PSK as an example. As discussed in \cite{r45}, we denote $\vec {OS}=s_k$ and $\vec {OA}=t \cdot s_k$, where $t>0$ is the objective to be maximized. $\vec {OB}$ is the received signal excluding noise, and based on \eqref{eq_1} $\vec {OB}$ is expressed as
\begin{equation}
\vec {OB} = {\bf h}_k^T {\bf Ws} = \lambda_k s_k,
\label{eq_2}
\end{equation}
where $\lambda_k$ is an introduced complex scalar that represents the interference effect on user $k$'s data symbol. Based on the geometry in Fig. 1 and \cite{r45}, the CI constraint is to locate the noiseless received signal $\vec {OB}$ within the constructive area, i.e, $|\vec {CB}| \le |\vec {CD}|$, which mathematically leads to 
\begin{equation}
\tan {\theta _{AB}} \le \tan {\theta _t} {\kern 5pt}
\Rightarrow {\kern 5pt} \left( {\lambda _k^\Re  - t} \right)\tan {\theta _t} \ge \left| {\lambda _k^\Im } \right|,
\label{eq_3}
\end{equation}
where $\lambda_k^\Re$ and $\lambda_k^\Im$ represent the real and imaginary part of $\lambda_k$, respectively. Throughout this paper, we focus on the non-strict phase-rotation CI, while the strict phase-rotation CI can be regarded as a special case by setting each $\lambda_k^\Im$ to zero \cite{r45}. Based on the constellation, we also obtain $\theta_t = \frac{\pi}{\cal M}$ for $\cal M$-PSK modulation. Accordingly, the optimization on CI precoding that maximizes the CI effect subject to the total available transmit power can be constructed as:
\begin{equation}
\begin{aligned}
&\mathcal{P}_1: {\kern 3pt} \mathop {\max }\limits_{{{\bf{W}}}} {\kern 3pt} t \\
&{\kern 0pt} s. t. {\kern 10pt} {{\bf{h}}_k^T}{\bf Ws} = {\lambda _k}{s_k}, {\kern 3pt} \forall k \in {\cal K} \\
&{\kern 22pt} \left( {\lambda _k^\Re  - t} \right)\tan {\theta _t} \ge \left| {\lambda _k^\Im } \right|, {\kern 3pt} \forall k \in {\cal K}\\
&{\kern 22pt} \left\| {{\bf{Ws}}} \right\|_2^2 \le {p_0}
\label{eq_4}
\end{aligned}
\end{equation}
where ${\cal K}=\left\{ {1,2,\cdots,K} \right\}$. In ${\cal P}_1$, we have enforced a symbol-level power constraint, since the interference exploitation precoding is dependent on the data symbol vector $\bf s$. ${\cal P}_1$ belongs to the second-order cone programming (SOCP) and can be solved via existing convex optimization tools.

\section{Interference Exploitation Precoding}
In this section, we analyze the interference exploitation precoding problem ${\cal P}_1$ based on the KKT conditions. Specifically, our derivations in this section and the numerical results in Section IV show that, by exploiting the information of both the channel and the data symbols, CI precoding is capable of spatially multiplexing more data streams than the number of transmit antennas at the BS simultaneously.

Following \cite{r18}, \cite{r45} and based on the observation that ${\bf Ws}$ can be viewed as a single vector in the formulation of ${\cal P}_1$, it is safe to assume that each ${\bf w}_i s_i$ is identical, which leads to a simpler power constraint in the subsequent analysis, given by
\begin{equation}
\sum\limits_{i = 1}^K {s_i^*{\bf{w}}_i^H{{\bf{w}}_i}{s_i}} \le \frac{p_0}{K}.
\label{eq_5}
\end{equation}
We then express ${\cal P}_1$ in a standard minimization form as
\begin{equation}
\begin{aligned}
&\mathcal{P}_2: {\kern 3pt} \mathop {\min }\limits_{{{\bf{W}}}} {\kern 3pt} -t \\
&{\kern 0pt} s. t. {\kern 10pt} {{\bf{h}}_k^T}\sum\limits_{i = 1}^K {{{\bf{w}}_i}{s_i}} - \lambda_k s_k=0, {\kern 3pt} \forall k \in {\cal K} \\
&{\kern 22pt}  \left| {\lambda _k^\Im } \right| - \left( {\lambda _k^\Re  - t} \right)\tan {\theta _t} \le 0, {\kern 3pt} \forall k \in {\cal K} \\
&{\kern 22pt} \sum\limits_{i = 1}^K {s_i^*{\bf{w}}_i^H{{\bf{w}}_i}{s_i}} \le \frac{p_0}{K}
\label{eq_6}
\end{aligned}
\end{equation}
and the Lagrangian of ${\cal P}_1$ is given by \cite{r27}
\begin{equation}
\begin{aligned}
{\cal L} \left( {{\bf W}, t, \delta_k, \tau_k, \delta_0} \right)=&  - t + \sum\limits_{k = 1}^K {{\delta _k}\left( {{{\bf{h}}_k}\sum\limits_{i = 1}^K {{{\bf{w}}_i}{s_i}}  - {\lambda _k}{s_k}} \right)} \\
&+ \sum\limits_{k = 1}^K {{\tau _k}\left[ {\left| {\lambda _k^\Im} \right| - \left( {\lambda _k^\Re - t} \right)\tan {\theta _t}} \right]} \\
& + {\delta _0}\left( {\sum\limits_{i = 1}^K {s_i^H{\bf{w}}_i^H{{\bf{w}}_i}{s_i}}  - \frac{p_0}{K}} \right),
\label{eq_7}
\end{aligned}
\end{equation}
where $\delta_k$, $\tau_k \ge 0$ and $\delta_0 \ge 0$ are the dual variables corresponding to each constraint of ${\cal P}_2$. Based on \eqref{eq_6} and the fact that $s_i s_i^H=1$, $\forall i \in {\cal K}$, the KKT conditions can be formulated as \cite{r27}
\begin{IEEEeqnarray}{rCl} 
\IEEEyesnumber
\frac{{\partial {\cal L}}}{{\partial t}} =  - 1 + \tan {\theta _t} \sum\limits_{k = 1}^K {{\tau _k}}  = 0 {\kern 30pt} \IEEEyessubnumber* \label{eq_8a} \\
\frac{{\partial {\cal L}}}{{\partial {{\bf{w}}_i}}} = \left( {\sum\limits_{k = 1}^K {{\delta _k} \cdot {{\bf{h}}_k}} } \right){s_i} + {\delta _0} \cdot {\bf{w}}_i^H = {\bf{0}}, {\kern 2pt} \forall i \in {\cal K} {\kern 30pt} \label{eq_8b} \\
{\delta _k}\left( {{{\bf{h}}_k}\sum\limits_{i = 1}^K {{{\bf{w}}_i}{s_i}}  - {\lambda _k}{s_k}} \right) = 0, {\kern 2pt} \forall k \in {\cal K}{\kern 30pt} \label{eq_8c} \\
{\tau _k}\left[ {\left| {\lambda _k^\Im} \right| - \left( {\lambda _k^\Re - t} \right)\tan {\theta _t}} \right] = 0, {\kern 2pt} \forall k \in {\cal K} {\kern 30pt} \label{eq_8d}\\
{\delta _0}\left( {\sum\limits_{i = 1}^K {s_i^H{\bf{w}}_i^H{{\bf{w}}_i}{s_i}}  - \frac{p_0}{K}} \right) = 0 {\kern 30pt} \label{eq_8e}
\end{IEEEeqnarray}
We first obtain that $\delta_0 \ne 0$ based on \eqref{eq_8b}, and accordingly we can express ${\bf w}_i^H$ as 
\begin{equation}
{\bf{w}}_i^H =  - \left( {\sum\limits_{k = 1}^K {\frac{{{\delta _k}}}{{{\delta _0}}} \cdot {{\bf{h}}_k}} } \right){s_i}, {\kern 3pt} \forall i \in {\cal K}.
\label{eq_9}
\end{equation}
By introducing a new variable $\upsilon_k=-\frac{{{\delta _k^H}}}{{{\delta _0}}}$, where we note that $\delta_k$ can be complex, we can further obtain the expression of ${\bf w}_i$, given by
\begin{equation}
{{\bf{w}}_i} = \left( {\sum\limits_{k = 1}^K {{\upsilon _k} \cdot {\bf{h}}_k^H} } \right)s_i^H, {\kern 3pt} \forall i \in {\cal K},
\label{eq_10}
\end{equation}
and we can further obtain the expression of each ${\bf w}_i s_i$ as
\begin{equation}
{{\bf{w}}_i}s_i = \sum\limits_{k = 1}^K {{\upsilon _k} \cdot {\bf{h}}_k^H},
\label{eq_11}
\end{equation}
which is constant for $\forall i \in {\cal K}$ and is consistent with our premise for the power constraint transformation in \eqref{eq_5}. Based on \eqref{eq_10}, we are now able to express the precoding matrix $\bf W$ as
\begin{equation}
\begin{aligned}
{\bf{W}} & = \left[ {{{\bf{w}}_1},{{\bf{w}}_2}, \cdots ,{{\bf{w}}_K}} \right] \\
& = \left( {\sum\limits_{k = 1}^K {{\upsilon_k} \cdot {\bf{h}}_k^H} } \right) \cdot \left[ {s_1^H,s_2^H, \cdots ,s_K^H} \right] \\
& = \left[ {{\bf{h}}_1^H,{\bf{h}}_2^H, \cdots ,{\bf{h}}_K^H} \right]{\left[ {{\upsilon_1},{\upsilon_2}, \cdots ,{\upsilon_K}} \right]^T}\left[ {s_1^H,s_2^H, \cdots ,s_K^H} \right] \\
& = {{\bf{H}}^H}{\bf{\Upsilon}}{{\bf{s}}^H},
\end{aligned}
\label{eq_12}
\end{equation}
Based on \eqref{eq_1} and \eqref{eq_2}, we can express the received signal vector excluding noise $\bf HWs$ as
\begin{equation}
{\bf HWs} = diag \left( \bf \Lambda \right) {\bf s},
\label{eq_13}
\end{equation}
where ${\bf \Lambda}=\left[ {\lambda_1,\lambda_2,\cdots,\lambda_K} \right]^T \in {\cal C}^{K \times 1}$ denotes the pre-scaling vector. By substituting \eqref{eq_12} into \eqref{eq_13}, we further obtain
\begin{equation}
\begin{aligned}
&{\bf{H}}{{\bf{H}}^H}{\bf \Upsilon} {{\bf{s}}^H}{\bf{s}} = diag\left( \bf \Lambda  \right){\bf{s}}\\
\Rightarrow & {\bf \Upsilon} = \frac{1}{K} \cdot \left( {{\bf{H}}{{\bf{H}}^H}} \right)^+ diag\left( \bf \Lambda  \right){\bf{s}},
\end{aligned}
\label{eq_14}
\end{equation}
where we note that based on the premise that $K > N_t$, the matrix ${\bf HH}^H$ is rank-deficient and the exact matrix inverse is inapplicable. Therefore, pseudo inverse has to be employed in \eqref{eq_14}, and subsequently we can obtain the optimal precoding structure as a function of the pre-scaling vector, given by
\begin{equation}
{\bf W}=\frac{1}{K} \cdot {\bf H}^H \left( {{\bf{H}}{{\bf{H}}^H}} \right)^+ diag\left( \bf \Lambda  \right){\bf{s}}{\bf s}^H.
\label{eq_15}
\end{equation}
Based on the fact that $\delta_0 \ne 0$, we obtain that the power constraint is strictly active. Similar to the analysis for the case of $K \le N_t$ in \cite{r45}, by substituting the expression of $\bf W$ into the power constraint $\left\| {{\bf{Ws}}} \right\|_2^2 = {p_0}$, one can similarly transform the power constraint on $\bf W$ into a power constraint on the pre-scaling vector $\bf \Lambda$, given by
\begin{equation}
\begin{aligned}
&\left\| {{\bf{Ws}}} \right\|_2^2 = {p_0} \\
\Rightarrow &{\kern 2pt} {{\bf{s}}^H}{{\bf{W}}^H}{\bf{Ws}} = {p_0}\\
\Rightarrow &{\kern 2pt} {{\bf{s}}^H}diag\left( {{{\bf \Lambda}^H}} \right){\left( {{\bf{H}}{{\bf{H}}^H}} \right)^ + }diag\left( \bf \Lambda  \right){\bf{s}} = {p_0}\\
\Rightarrow &{\kern 2pt} {{\bf \Lambda} ^H}\underbrace {diag\left( {{{\bf{s}}^H}} \right){{\left( {{\bf{H}}{{\bf{H}}^H}} \right)}^ + }diag\left( {\bf{s}} \right)}_{\bf{P}}{\bf \Lambda} = {p_0}.
\end{aligned}
\label{eq_16}
\end{equation}
Then, one can follow a similar approach in \cite{r45} to obtain an equivalent QP formulation. 

However, in the case of $K > N_t$ considered in this paper, we note that following the above procedure and using \eqref{eq_16} will only lead to an erroneous solution for CI precoding, which is due to the fact that the inclusion of the pseudo inverse does not guarantee equality to the original constraint. To be more specific, let's first consider the conventional case of $K \le N_t$, where the optimal precoding structure is given by \cite{r45}
\begin{equation}
{\bf W}=\frac{1}{K} \cdot {\bf H}^H \left( {{\bf{H}}{{\bf{H}}^H}} \right)^{-1} diag\left( \bf \Lambda  \right){\bf{s}}{\bf s}^H.
\label{eq_17}
\end{equation}
In this case, by substituting the expression of $\bf W$ in \eqref{eq_17} into \eqref{eq_13}, we obtain
\begin{equation}
\begin{aligned}
&{\kern 2pt} {\bf HWs} = diag \left( \bf \Lambda \right) {\bf s} \\
\Rightarrow &{\kern 2pt} {\bf{H}}\left[ {\frac{1}{K} \cdot {{\bf{H}}^H}{{\left( {{\bf{H}}{{\bf{H}}^H}} \right)}^{ - 1}}diag\left( \bf \Lambda  \right){\bf{s}}{{\bf{s}}^H}} \right]{\bf{s}} = diag\left( \bf \Lambda  \right){\bf{s}}\\
\Rightarrow &{\kern 2pt} diag\left( \bf \Lambda  \right){\bf{s}} = diag\left( \bf \Lambda  \right){\bf{s}},
\end{aligned}
\label{eq_18}
\end{equation}
which is always true. This in fact means that the pre-scaling constraint in \eqref{eq_13} is already included in the power constraint implicitly, for the case of $K \le N_t$. On the contrary, in the considered scenario of $K > N_t$ in this paper where the pseudo inverse is included, the above equality will not hold, and simply following a similar approach to the case of $K \le N_t$ in \cite{r45} will lead to invalid and erroneous solutions. Therefore, an additional constraint is required to guarantee that the inclusion of pseudo inverse still meets the pre-scaling requirement, given by
\begin{equation}
\begin{aligned}
&{\kern 2pt} {\bf HWs} = diag \left( \bf \Lambda \right) {\bf s} \\
\Rightarrow &{\kern 2pt} {\bf{H}}\left[ {\frac{1}{K} \cdot {{\bf{H}}^H}{{\left( {{\bf{H}}{{\bf{H}}^H}} \right)}^ + }diag\left( \bf \Lambda  \right){\bf{s}}{{\bf{s}}^H}} \right]{\bf{s}} = diag\left( \bf \Lambda \right){\bf{s}}\\
\Rightarrow &{\kern 2pt} {\bf{H}}{{\bf{H}}^H}{\left( {{\bf{H}}{{\bf{H}}^H}} \right)^ + }diag\left( \bf \Lambda \right){\bf{s}} = diag\left( \bf \Lambda  \right){\bf{s}}\\
\Rightarrow & \left[ {{\bf{H}}{{\bf{H}}^H}{{\left( {{\bf{H}}{{\bf{H}}^H}} \right)}^ + } - {\bf{I}}} \right]diag\left( \bf \Lambda  \right){\bf{s}} = {\bf{0}} \\
\Rightarrow & \underbrace {\left[ {{\bf{H}}{{\bf{H}}^H}{{\left( {{\bf{H}}{{\bf{H}}^H}} \right)}^ + } - {\bf{I}}} \right]diag\left( {\bf{s}} \right)}_{\bf{T}}{\bf \Lambda} = {\bf{0}}
\end{aligned}
\label{eq_19}
\end{equation}
Based on \eqref{eq_19}, first we observe that ${\bf \Lambda}={\bf 0}$ is obviously not a valid solution to the original CI precoding. Accordingly, this additional constraint is equivalent to finding non-zero solutions to the linear equation set ${\bf T \Lambda}={\bf 0}$. Noting that both $\bf T$ and $\bf \Lambda$ are complex, we first transform them into their real equivalence, given by
\begin{equation}
{{\bf{T}}_{\bf{E}}} = \left[ {\begin{array}{*{20}{c}}
{\Re \left( {\bf{T}} \right)}&{ - \Im \left( {\bf{T}} \right)}\\
{\Im \left( {\bf{T}} \right)}&{\Re \left( {\bf{T}} \right)}
\end{array}} \right], {\kern 3pt} {\bf \Lambda _E} = \left[ {\begin{array}{*{20}{c}}
{\Re \left( \bf \Lambda  \right)}\\
{\Im \left( \bf \Lambda  \right)}
\end{array}} \right],
\label{eq_20}
\end{equation}
and we further express the singular value decomposition (SVD) of $\bf T_E$ as
\begin{equation}
{{\bf{T}}_{\bf{E}}} = {\bf{S}}{\bf \Sigma} {{\bf{\hat D}}^H},
\label{eq_21}
\end{equation}
where ${\bf{\hat D}} = \left[ {{{\bf \hat{d}}_1},{{\bf \hat{d}}_2}, \cdots ,{{\bf \hat{d}}_{2K}}} \right]$ is the right singular matrix that consists of right singular vectors. Based on the linear algebra theory \cite{r47}, the non-zero solution $\bf \Lambda_E$ is therefore in the null space of $\bf T_E$, which can be expressed as a linear combination of the right singular vectors that correspond to zero singular values, given by
\begin{equation}
{\bf \Lambda _E} = \sum\limits_{n = 1}^{2K - {\rm{rank}}\left\{ {{{\bf{T}}_{\bf{E}}}} \right\}} {{\beta _n} \cdot {{{\bf{\hat d}}}_{{\rm{rank}}\left\{ {{{\bf{T}}_{\bf{E}}}} \right\} + n}}}  = {\bf{D}}{\bm \beta},
\label{eq_22}
\end{equation}
where each $\beta_n$ is real and ${\bm \beta} = {\left[ {{\beta _1},{\beta _2}, \cdots ,{\beta _{2K - {\rm{rank}}\left\{ {{{\bf{T}}_{\bf{E}}}} \right\}}}} \right]^T}$. $\bf D$ consists of right singular vectors corresponding to zero singular values, given by
\begin{equation}
\begin{aligned}
{\bf{D}} &= \left[ {{{{\bf{\hat d}}}_{{\rm{rank}}\left\{ {{{\bf{T}}_{\bf{E}}}} \right\} + 1}},{{{\bf{\hat d}}}_{{\rm{rank}}\left\{ {{{\bf{T}}_{\bf{E}}}} \right\} + 2}}, \cdots ,{{{\bf{\hat d}}}_{2K}}} \right]\\
& = \left[ {{{\bf{d}}_1^T},{{\bf{d}}_2^T}, \cdots ,{{\bf{d}}_{2K}^T}} \right]^T,
\end{aligned}
\label{eq_23}
\end{equation}
where each ${\bf d}_k^T$ represents the $k$-th row of $\bf D$. Subsequently, we expand the left-hand side of \eqref{eq_16} into its real equivalence, given by
\begin{equation}
{\bf \Lambda} _{\bf E}^T\underbrace {\left[ {\begin{array}{*{20}{c}}
{\Re \left( {\bf{P}} \right)}&{ - \Im \left( {\bf{P}} \right)}\\
{\Im \left( {\bf{P}} \right)}&{\Re \left( {\bf{P}} \right)}
\end{array}} \right]}_{{{\bf{P}}_{\bf{E}}}}{\bf \Lambda _E} = {p_0} {\kern 2pt}
\Rightarrow {\kern 2pt} {{\bm \beta}^T}\underbrace {{{\bf{D}}^T}{{\bf{P}}_{\bf{E}}}{\bf{D}}}_{{{\bf{Q}}_{\bf{E}}}}{\bm \beta}  = {p_0},
\label{eq_24}
\end{equation}
which is the valid power constraint for the case of $K >N_t$ considered in this paper, and we further note that $\bf Q_E$ is symmetric.

\newcounter{mytempeqncnt}
\begin{figure*}
\normalsize
\setcounter{mytempeqncnt}{\value{equation}}
\setcounter{equation}{33}

\begin{equation}
{\bf{W}} = \frac{1}{K} \cdot {{\bf{H}}^H}{\left( {{\bf{H}}{{\bf{H}}^H}} \right)^ + }diag\left\{ {\sqrt {\frac{{{p_0}}}{{{{\bf{u}}^T}{\bf{SDQ}}_{\bf{E}}^{ - 1}{{\bf{D}}^T}{{\bf{S}}^T}{\bf{u}}}}}  \cdot {\bf{UDQ}}_{\bf{E}}^{ - 1}{{\bf{D}}^T}{{\bf{S}}^T}{\bf{u}}} \right\}{\bf{s}}{{\bf{s}}^H}
\label{eq_34}
\end{equation}

\setcounter{equation}{\value{mytempeqncnt}}
\hrulefill
\vspace*{4pt}
\end{figure*}

Based on the above analysis, we can now formulate an equivalent optimization on the weight vector $\bm \beta$, given by
\begin{equation}
\begin{aligned}
&\mathcal{P}_3: {\kern 3pt} \mathop {\min }\limits_{\bm \beta} {\kern 3pt} -t \\
&{\kern 0pt} s. t. {\kern 11pt} {\bm \beta}^T {\bf Q_E} {\bm \beta}- p_0 =0\\
&{\kern 23pt}  \frac{{{\bf{d}}_{k + K}^T {\bm \beta} }}{{\tan {\theta _t}}} + t - {\bf{d}}_k^T{\bm \beta} \le 0, {\kern 3pt} \forall k \in {\cal K}\\
&{\kern 22pt}  -\frac{{{\bf{d}}_{k + K}^T{\bm \beta} }}{{\tan {\theta _t}}} + t - {\bf{d}}_k^T {\bm \beta} \le 0, {\kern 3pt} \forall k \in {\cal K}
\label{eq_25}
\end{aligned}
\end{equation}
where we have transformed the CI constraint with the absolute value into two separate constraints. The Lagrangian of ${\cal P}_3$ is constructed as 
\begin{equation}
\begin{aligned}
{\cal L} \left( {{\bm \beta}, t, \alpha_0, \mu_k, \nu_k} \right) = &  - t + \alpha_0 \left({{{\bm \beta} ^T}{{\bf{Q}}_{\bf{E}}}{\bm \beta}-p_0}\right) \\
& + \sum\limits_{k = 1}^K {{\mu _k}\left( {\frac{{{\bf{d}}_{k + K}^T{\bm \beta} }}{{\tan {\theta _t}}} + t - {\bf{d}}_k^T {\bm \beta} } \right)} \\
& + \sum\limits_{k = 1}^K {{\nu _k}\left( { - \frac{{{\bf{d}}_{k + K}^T {\bm \beta} }}{{\tan {\theta _t}}} + t - {\bf{d}}_k^T {\bm \beta} } \right)},
\end{aligned}
\label{eq_26}
\end{equation}
where $\alpha_0$, $\mu_k \ge 0$ and $\nu_k \ge 0$ are the corresponding dual variables. By defining
\begin{equation}
{\bf{u}} = {\left[ {{\mu _1}, \cdots ,{\mu _K},{\nu _1}, \cdots ,{\nu _K}} \right]^T}, {\kern 3pt}
{\bf{S}} = \left[ {\begin{array}{*{20}{c}}
{\bf{I}}&{ - \frac{1}{{\tan {\theta _t}}} \cdot {\bf{I}}}\\
{\bf{I}}&{\frac{1}{{\tan {\theta _t}}} \cdot {\bf{I}}}
\end{array}} \right],
\label{eq_27}
\end{equation}
where ${\bf u} \in {\cal R}^{2K \times 1}$ and ${\bf S} \in {\cal R}^{2K \times 2K}$. The Lagrangian of ${\cal P}_3$ can be further simplified into
\begin{equation}
{\cal L} \left( {{\bm \beta}, t, \alpha_0, {\bf u}} \right) =\left( {{{\bf{1}}^T}{\bf{u}} - 1} \right)t + {\alpha _0} \cdot {{\bm \beta} ^T}{{\bf{Q}}_{\bf{E}}}{\bm \beta}  - {{\bf{u}}^T}{\bf{SD}}{\bm \beta}  - {\alpha _0}{p_0},
\label{eq_28}
\end{equation}
based on which the KKT conditions for ${\cal P}_3$ are given by
\begin{IEEEeqnarray}{rCl} 
\IEEEyesnumber
\frac{{\partial {\cal L}}}{{\partial t}} = {{\bf{1}}^T}{\bf{u}} - 1 = 0  \IEEEyessubnumber* \label{eq_29a} \\
\frac{{\partial {\cal L}}}{{\partial {\bm \beta} }} = 2{\alpha _0} \cdot {\bf Q_E}{\bm \beta} - {\bf D}^T{{\bf{S}}^T}{\bf{u}} = {\bf{0}} \label{eq_29b} \\
{\bm \beta}^T {\bf Q_E} {\bm \beta}- p_0 =0 \label{eq_29c} \\
{\mu _k}\left( { \frac{{{\bf{d}}_{k + K}^T {\bm \beta} }}{{\tan {\theta _t}}} + t - {\bf{d}}_k^T{\bm \beta} \le 0 } \right) = 0, {\kern 3pt} \forall k \in {\cal K} \label{eq_29d} \\
{\nu _k}\left( { -\frac{{{\bf{d}}_{k + K}^T {\bm \beta} }}{{\tan {\theta _t}}} + t - {\bf{d}}_k^T{\bm \beta} \le 0 } \right) = 0, {\kern 3pt} \forall k \in {\cal K} \label{eq_29e}
\end{IEEEeqnarray}
Based on \eqref{eq_29b} we obtain $\alpha_0 \ne 0$ and the expression of $\bm \beta$ as a function of $\bf u$, given by
\begin{equation}
{\bm \beta}  = \frac{1}{{2{\alpha _0}}} \cdot {\bf{Q}}_{\bf{E}}^{ - 1}{{\bf{D}}^T}{{\bf{S}}^T}{\bf{u}}.
\label{eq_30}
\end{equation}
By substituting the expression of $\bm \beta$ into \eqref{eq_24}, we can further obtain the expression of $\alpha_0$, given by
\begin{equation}
\begin{aligned}
&{\left( {\frac{1}{{2{\alpha _0}}} \cdot {\bf{Q}}_{\bf{E}}^{ - 1}{{\bf{D}}^T}{{\bf{S}}^T}{\bf{u}}} \right)^T}{{\bf{Q}}_{\bf{E}}}\left( {\frac{1}{{2{\alpha _0}}} \cdot {\bf{Q}}_{\bf{E}}^{ - 1}{{\bf{D}}^T}{{\bf{S}}^T}{\bf{u}}} \right) = {p_0}\\ 
\Rightarrow & {\kern 2pt} {\alpha _0} = \sqrt {\frac{{{{\bf{u}}^T}{\bf{SDQ}}_{\bf{E}}^{ - 1}{{\bf{D}}^T}{{\bf{S}}^T}{\bf{u}}}}{{4{p_0}}}} 
\end{aligned}
\label{eq_31}
\end{equation}
For ${\cal P}_3$, it is easy to verify that the Slater's condition is satisfied, and therefore we consider the dual problem of ${\cal P}_3$:
\begin{equation}
\begin{aligned}
{\cal D} =& \mathop {\max }\limits_{{\alpha _0},{\bf{u}}} \mathop {\min }\limits_{{\bm \beta} ,t} {\cal L}\left( {\beta ,t,{\alpha _0},{\bf{u}}} \right)\\
=&\mathop {\max }\limits_{{\alpha _0},{\bf{u}}} {\alpha _0}{\left( {\frac{1}{{2{\alpha _0}}}  {\bf{Q}}_{\bf{E}}^{ - 1}{{\bf{D}}^T}{{\bf{S}}^T}{\bf{u}}} \right)^T}{{\bf{Q}}_{\bf{E}}}\left( {\frac{1}{{2{\alpha _0}}} {\bf{Q}}_{\bf{E}}^{ - 1}{{\bf{D}}^T}{{\bf{S}}^T}{\bf{u}}} \right)\\
& {\kern 20pt} - {{\bf{u}}^T}{\bf{SD}}\left( {\frac{1}{{2{\alpha _0}}} {\bf{Q}}_{\bf{E}}^{ - 1}{{\bf{D}}^T}{{\bf{S}}^T}{\bf{u}}} \right) - {\alpha _0}{p_0}\\
=&\mathop {\max }\limits_{{\alpha _0},{\bf{u}}}  \frac{1}{{4{\alpha _0}}} \cdot {{\bf{u}}^T}{\bf{SDQ}}_{\bf{E}}^{ - 1}{{\bf{D}}^T}{{\bf{S}}^T}{\bf{u}} - {\alpha _0}{p_0}\\
=&\mathop {\max }\limits_{{\bf{u}}} - \frac{{{{\bf{u}}^T}{\bf{SDQ}}_{\bf{E}}^{ - 1}{{\bf{D}}^T}{{\bf{S}}^T}{\bf{u}}}}{{4\sqrt {\frac{{{{\bf{u}}^T}{\bf{SDQ}}_{\bf{E}}^{ - 1}{{\bf{D}}^T}{{\bf{S}}^T}{\bf{u}}}}{{4{p_0}}}} }} - p_0\sqrt {\frac{{{{\bf{u}}^T}{\bf{SDQ}}_{\bf{E}}^{ - 1}{{\bf{D}}^T}{{\bf{S}}^T}{\bf{u}}}}{{4{p_0}}}} \\
=&\mathop {\max }\limits_{{\bf{u}}} - \sqrt {{p_0} \cdot {{\bf{u}}^T}{\bf{SDQ}}_{\bf{E}}^{ - 1}{{\bf{D}}^T}{{\bf{S}}^T}{\bf{u}}}.
\end{aligned}
\label{eq_32}
\end{equation}
Based on the monotonicity of the square root function, the maximization on the dual in \eqref{eq_32} is equivalent to the following optimization:
\begin{equation}
\begin{aligned}
&\mathcal{P}_4: {\kern 3pt} \mathop {\min }\limits_{\bf u} {\kern 3pt} {{{\bf{u}}^T} \left({{\bf{SDQ}}_{\bf{E}}^{ - 1}{{\bf{D}}^T}{{\bf{S}}^T}}\right) {\bf{u}}} \\
&{\kern 0pt} s. t. {\kern 10pt}  {{\bf{1}}^T}{\bf{u}} - 1 = 0\\
&{\kern 24pt}  u_k \ge 0, {\kern 3pt} \forall k \in \hat {\cal K}
\label{eq_33}
\end{aligned}
\end{equation}
where the first constraint is from \eqref{eq_29a}, $\hat {\cal K}=\left\{ {1,2,\cdots,2K} \right\}$, and $u_k$ is the $k$-th entry in $\bf u$.

Based on our above transformations and analysis, we have shown that the original CI precoding problem ${\cal P}_1$ can be equivalently solved by a simplified optimization ${\cal P}_4$. Moreover, based on the expression of $\bf \Lambda_E$ in \eqref{eq_22}, $\bm \beta$ in \eqref{eq_30}, and $\alpha_0$ in \eqref{eq_31}, we can express the optimal precoding matrix $\bf W$ as a function of the dual vector $\bf u$ in a closed form, which is shown in \eqref{eq_34} at the top of this page, where ${\bf U}=\left[ {\begin{array}{*{20}{c}}{\bf{I}}&{j \cdot {\bf{I}}}\end{array}} \right]$ transforms the expanded pre-scaling vector ${\bf \Lambda_E}$ into its complex equivalence $\bf \Lambda$.

Compared to the original CI precoding optimization in ${\cal P}_1$ which is a SOCP optimization, it is observed that ${\cal P}_4$ is a standard QP optimization over a simplex. It has been shown in the literature that such a QP formulation can be more efficiently solved than the SOCP formulation using the simplex or interior-point methods \cite{r29}, \cite{r44}, and the iterative closed-form algorithm proposed in \cite{r45} can also be directly employed to solve ${\cal P}_4$ with reduced computational costs.

\subsection{Condition for Spatially Multiplexing $K>N_t$ Streams}
Based on the above, we can also obtain the expression of $t^*$ when the optimality of ${\cal P}_3$ is achieved, given by
\setcounter{equation}{34}
\begin{equation}
{t^*} = \mathop {\min }\limits_k \left\{ {d_k^T{\bm \beta}  - \frac{{d_{k + K}^T {\bm \beta} }}{{\tan {\theta _t}}}, {\kern 3pt} d_k^T {\bm \beta}  + \frac{{d_{k + K}^T {\bm \beta} }}{{\tan {\theta _t}}}} \right\}, {\kern 3pt} \forall k \in \hat {\cal K}.
\label{eq_35}
\end{equation}
If $t^* > 0$, we obtain a valid pre-scaling vector $\bf \Lambda$ and a corresponding valid precoding matrix $\bf W$. Otherwise if $t^* \le 0$, it means that the data symbols will be scaled and rotated to other three quarters of the constellation, which only leads to erroneous detection. Accordingly, whether the obtained
\begin{equation}
\mathop {\min }\limits_k \left\{ {d_k^T{\bm \beta}  - \frac{{d_{k + K}^T {\bm \beta} }}{{\tan {\theta _t}}}, {\kern 3pt} d_k^T {\bm \beta}  + \frac{{d_{k + K}^T {\bm \beta} }}{{\tan {\theta _t}}}} \right\} > 0
\label{eq_36}
\end{equation}
is the condition under which multiplexing $K > N_t$ streams is feasible.

\section{Numerical Results}
Numerical results are presented based on Monte Carlo simulations in this section. We assume the total transmit power $p_0=1$, and define the transmit SNR as $\rho={1 \mathord{\left/{\vphantom {1 {{\sigma ^2}}}} \right.\kern-\nulldelimiterspace} {{\sigma ^2}}}$. Since ZF precoding and SINR balancing precoding are not feasible for the scenario of $K > N_t$, we compare our QP-based CI precoding with closed-form RZF precoding and traditional SOCP-based CI precoding. Both QPSK and 8PSK modulations are considered in the simulations.

\begin{figure}[!b]
\centering
\includegraphics[scale=0.35]{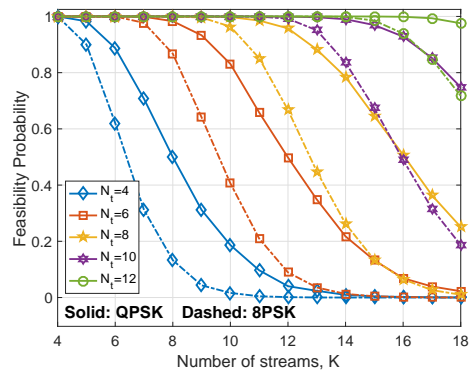}
\caption{Feasibility v.s. number of streams $K$}
\end{figure}

Before we present the bit error rate (BER) performance, we first depict the feasibility probability with respect to the number of streams $K$ in Fig. 2, where the number of transmit antennas $N_t$ varies from $N_t=4$ to $N_t=12$. Generally, for a specific feasibility target, we observe that a larger number of transmit antennas at the BS can support more streams than that of transmit antennas, i.e., a larger $N_t$ leads to a larger $\left( {K-N_t} \right)$. Specifically when $N_t=12$ for QPSK, CI precoding is able to support $K=18$ streams simultaneously with a feasibility probability higher than 95\%. For the following BER results in Fig. 3 and 4, RZF precoding is employed instead when CI precoding is not feasible in the simulations.

\begin{figure}[h]
\begin{centering}
\subfloat[$K=9$, $N_t=8$]
{\begin{centering}
\includegraphics[width=4cm]{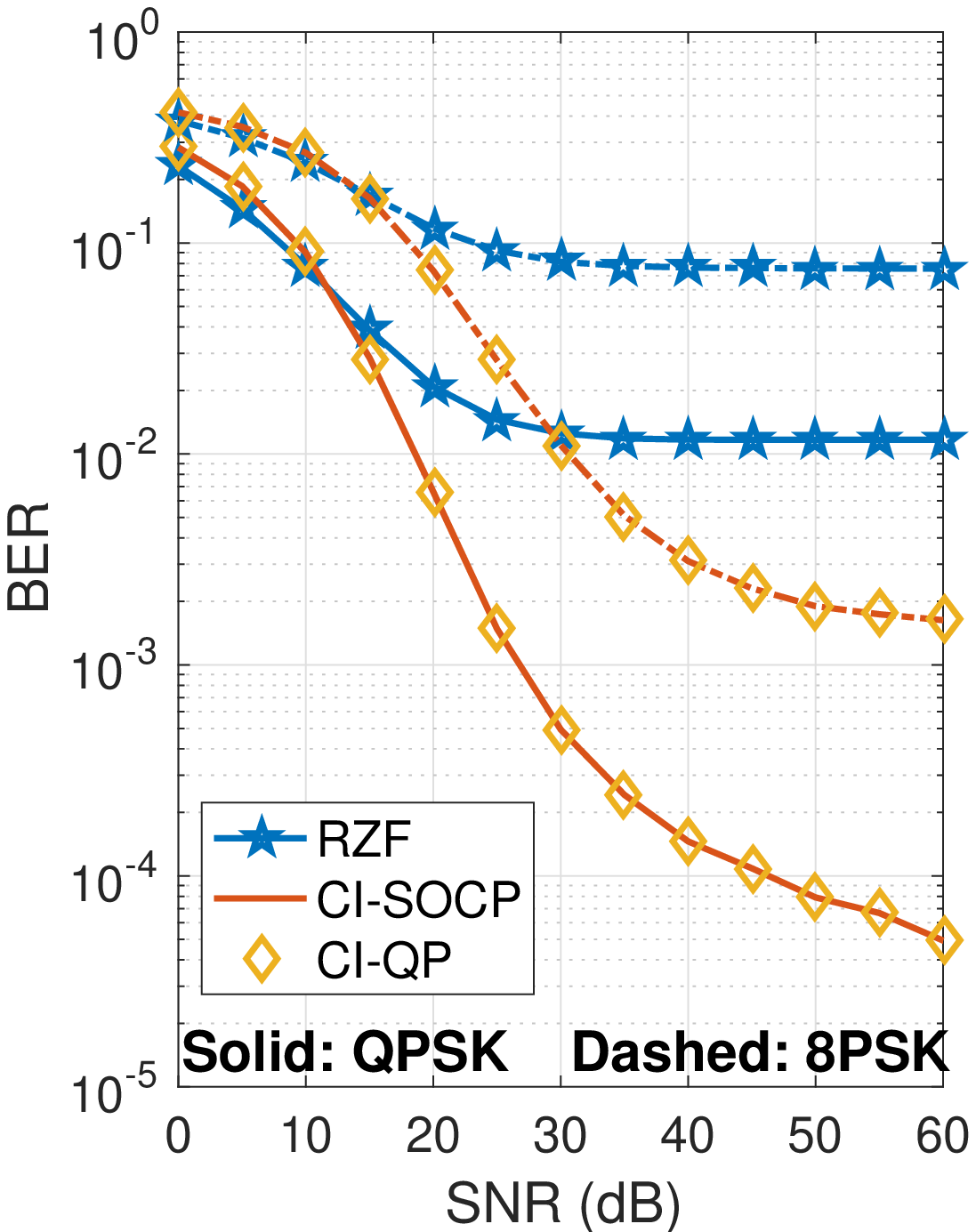}
\par\end{centering}
}\subfloat[$K=10$, $N_t=8$]
{\begin{centering}
\includegraphics[width=4cm]{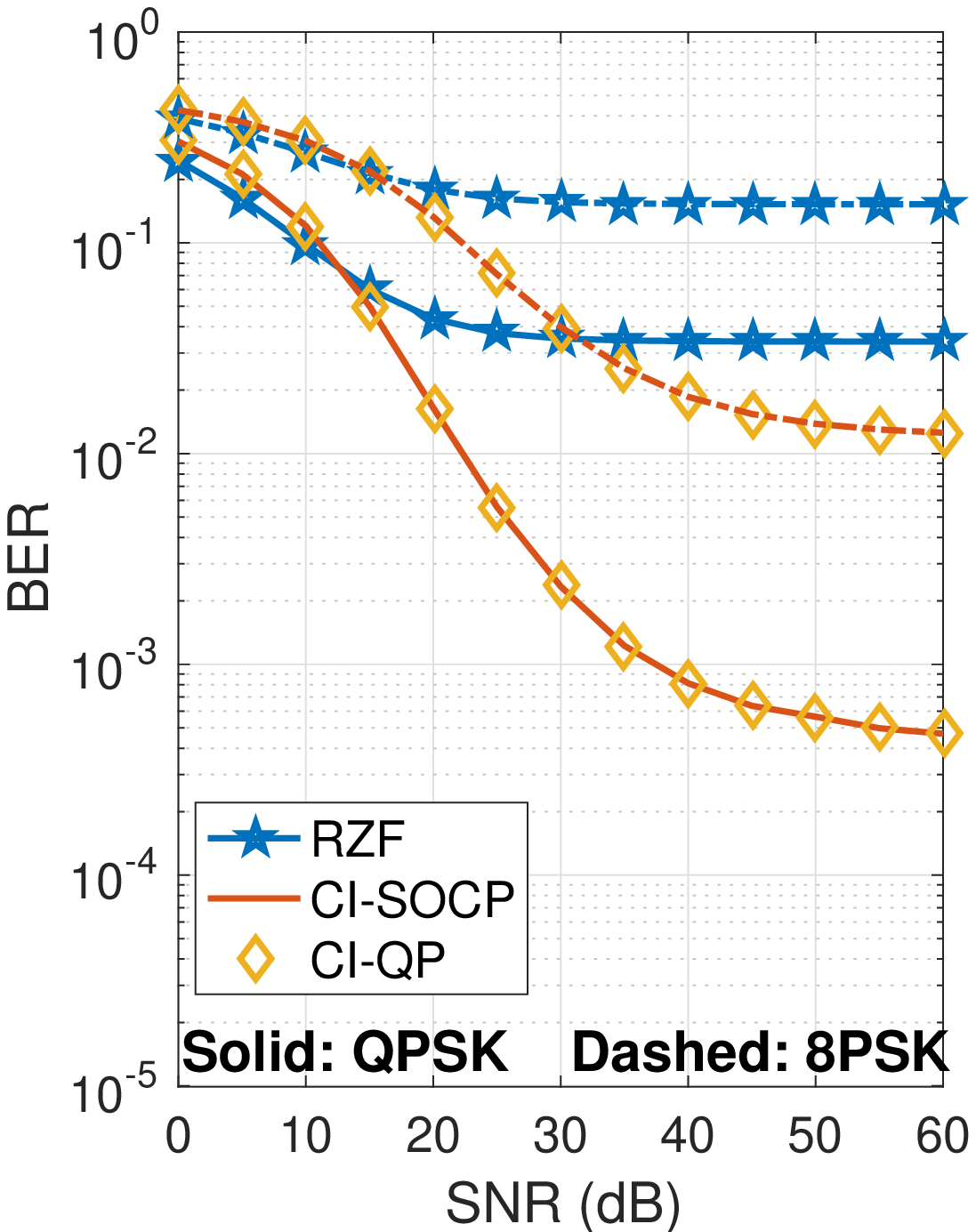}
\par\end{centering}
}
\par\end{centering}
\caption{\label{fig:PF}BER v.s. transmit SNR, $N_t=8$}
\end{figure}

The BER results of CI precoding are presented in Fig. 3 with respect to the transmit SNR $\rho$, where we consider two scenarios $K=9$, $N_t=8$ and $K=10$, $N_t=8$ for both QPSK and 8PSK modulation. When $K>N_t$, traditional ZF precoding and SINR balancing precoding are inapplicable, and therefore we compare with RZF precoding only. In Fig. 3, compared to RZF precoding where an error floor is observed, CI precoding achieves a significant performance gain in the medium-to-high SNR regime, which validates the superiority of CI precoding over traditional RZF precoding.

\begin{figure}[!t]
\centering
\includegraphics[scale=0.35]{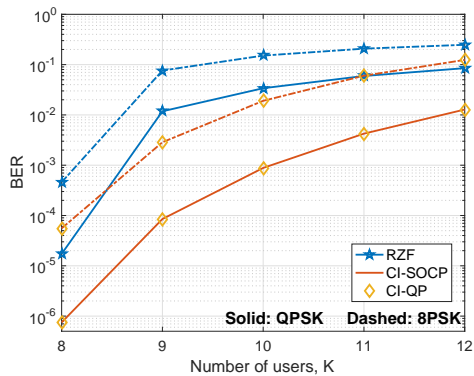}
\caption{BER v.s. number of streams $K$, $N_t=8$, SNR=40dB}
\end{figure}

In Fig. 4, we show the BER results of CI precoding with an increasing number of users $K$ for both QPSK and 8PSK modulation, where $N_t=8$ and $\rho=40$dB. For both modulations considered in Fig. 4, we observe a significant gain of CI precoding over traditional RZF precoding, when $K>N_t$. The performance gains become less significant when $K$ increases, which is due to a lower feasibility probability for CI precoding, as observed in Fig. 2.

\section{Conclusion}
In this paper, the interference exploitation precoding for the scenario where the BS serves a larger number of users than that of the transmit antennas is studied. By analyzing the optimization problem with KKT conditions and by formulating the dual problem, we obtain the closed-form optimal precoding matrix as a function of the dual vector, as well a QP optimization that efficiently obtains the optimal dual vector. Numerical results validate the optimality of the closed-form precoding matrix, and reveal significant performance gains of interference exploitation precoding over traditional RZF precoding.

\bibliographystyle{IEEEtran}
\bibliography{refs.bib}

\end{document}